# The first confirmed superoutburst of the SU UMa type dwarf nova SDSS J083931.35+282824.0

Jeremy Shears, Enrique de Miguel, George Roberts, Donald F. Collins, Gordon Myers and Tut Campbell

## Abstract

We report unfiltered CCD photometry of the first confirmed superoutburst of the recently discovered dwarf nova, SDSS J083931.35+282824.0 in April 2010. From a quiescence magnitude of ~19.8 it rose to 14.0, an outburst amplitude of at least 5.8 magnitudes. Only the plateau phase of the outburst was observed during which superhumps with peak-to-peak amplitude of up to 0.28 magnitudes were present, confirming this to be an SU UMa type dwarf nova. The mean superhump period was $P_{sh}$ = 0.07836(2) during the first 3 days and this subsequently decreased to 0.07800(3) d. Analysis of the data revealed tentative evidence for an orbital period $P_{orb}$ = 0.07531(25) d. The fractional superhump period excess was ε = 0.039(6), which is consistent with other dwarf novae of similar orbital period.

## Introduction

SU UMa type systems are dwarf novae which occasionally undergo *superoutbursts* during which *superhumps* are observed in the light curve. These are modulations in the light curve with a period a few percent longer than the orbital period. The systems are semi-detached binaries in which a white dwarf primary accretes material from a secondary star via Roche lobe overflow. The secondary is usually a late-type main-sequence star. Material from the secondary passes through an accretion disc before settling on the surface of the white dwarf. As material builds up in the disc, a thermal instability is triggered that drives the disc into a hotter, brighter state causing an outburst in which the star apparently brightens by several magnitudes (1). In SU UMa systems, the disc becomes eccentric because of a 3:1 resonance between the secondary star orbit and the motion of matter in the outer accretion disc. For a more detailed review of SU UMa dwarf novae and superhumps, the reader is directed to reference (1).

SDSS J083931.35+282824.0 was identified as a dwarf nova during the course of a search for such objects in data from the Sloan Digital Sky Survey (SDSS) by Szkody *et al.* (2). SDSS lists the object as having g = 20.22 and r = 20.17. We examined data from Catalina Real-Time Transient Survey (CRTS) (3) and found a mean quiescence magnitude of V=19.8 with a standard deviation of 1.4 magnitudes; however in many cases, the object was below the detection limit.

The outburst discussed in the paper was detected by K. Itagaki on 2010 Apr 8.585 at magnitude 14.0 (4). Following the announcement, the authors conducted an intensive campaign of time-resolved photometry facilitated by the Center for Backyard Astrophysics (5) (CBA), a global network of small telescopes dedicated to photometry of cataclysmic variable stars.

**Photometry and analysis**

The authors obtained 50 hours of unfiltered photometry during the outburst, using the instrumentation shown in Table 1 and according to the observation log in Table 2. Images were dark-subtracted and flat-fielded prior to being measured using differential aperture photometry relative to either NOMAD1 1184-0173155 (V=11.412) or NOMAD1 1184-0173124 (V=13.370) using magnitudes from the Naval Observatory Merged Astrometric Dataset (NOMAD) (6). Given that each observer used slightly different instrumentation, including CCD cameras with different spectral responses, small systematic differences are likely to exist between observers. Where overlapping datasets were obtained, we aligned measurements by different observers by experiment. Adjustments of up to 0.09 magnitudes were made. However, given that the aim of the time resolved photometry was to investigate periodic variations in the light curve, we consider this not to be a significant disadvantage. Heliocentric corrections were applied to all data.

**Outburst light curve**

The overall light curve of the outburst is shown in the top panel of Figure 1. The red data point is Itagaki's detection and the rest of the data are from the authors. Thus at its brightest the star was magnitude 14.0 and, taking the CRTS mean quiescence magnitude of 19.8, the outburst amplitude was at least 5.8 magnitudes. The 6 days during which the star was observed corresponded to the plateau phase of the outburst. During this period there was a gradual fading at a mean rate of 0.11mag/d, although it appeared to flatten out towards the end. Unfortunately, the decline to quiescence was not observed, although it had evidently reached quiescence by JD 2455309, some 14 days after detection, when it was fainter than magnitude 16.9.

**Measurement of the superhump period**

In Figure 2 we plot expanded views of the longer time series photometry runs, where each panel shows 2 days of data drawn to the same scale. This clearly shows the presence of regular modulations throughout the period of observation which we interpret as superhumps. The presence of superhumps is diagnostic that SDSS J083931.35+282824.0 is a member of the SU UMa family of dwarf novae, making this the first confirmed superoutburst of the star. The fact that fully developed superhumps were present on the first night of observing suggests that the superoutburst was already well established.

To study the superhump behaviour, we first extracted the times of each sufficiently well-defined superhump maximum by fitting a quadratic function to the peaks of the individual light curves. Times of 26 superhump maxima were found and are listed in Table 3 (although we note that the errors are likely to be underestimates). An analysis of the times of maximum for cycles 0 to 31 (JD 2455295 to 2455297), assuming a linear fit, allowed us to obtain the following linear superhump maximum ephemeris:

$$HJD_{max} = 2455295.3524(3) + 0.07836(2) \times E \qquad \text{Equation 1}$$

We used an unweighted fit since our timing errors are likely to be underestimates.

This gives the mean superhump period for the first three nights of the superoutburst as $P_{sh}$ = 0.07836(2) d. The O–C residuals for the superhump maxima for the complete outburst relative to the ephemeris are shown in the middle panel in Figure 1.

**Superhump evolution**

The O-C diagram shows that the superhump period appeared to be constant from the beginning of the observations until JD 2455298 (cycle 42). At around this time the period changed and a new slightly shorter period of 0.07800(3) d was present until the end of observations which is indicated by the blue dotted line if the O-C plot in Figure 1.

Kato *et al.* (7) studied the superhump period evolution in a large number of SU UMa systems and found that many superoutbursts appeared to show three distinct stages: an early evolutionary stage (A) with longer superhump period, a middle stage (B) during which systems with $P_{orb}$ < 0.08 d have a positive period derivative, and a final stage (C) with a shorter $P_{sh}$. We have attempted to interpret the O-C diagram of SDSS J083931.35+282824.0 in terms of Kato's model. It is possible that the interval from the detection of the outburst until JD 2455298 corresponds to stage B and that we observed a transition to stage C at JD 2455298. Fitting a quadratic function to the data from the beginning to JD 2455298 gave $dP_{sh}/dt$ = +4.1(9) x $10^{-4}$, the positive period derivative being consistent with stage B, but the residuals were only marginally improved relative to the linear fit. The main problem is that there are rather few data points during the possible stage B and C regimes to be certain of this classification.

Kato et al. (8) also analysed the same outburst of SDSS J083931.35+282824.0 as reported in this paper, although they included rather few superhump maxima in their analysis, and found a possible stage B to C transition. They also measured a mean superhump period $P_{sh}$ = 0.078423(7) d for data in the interval JD 2455295 to 2455298, which is very slightly longer than our value of $P_{sh}$ from Equation 1.

The peak-to-peak amplitude of the superhumps gradually decreased during the outburst from 0.28 magnitudes at the beginning to 0.16 magnitudes on the last night of observation (Figure 1, bottom panel and Table 3). As is sometimes observed in superoutbursts, there was considerable variation in amplitude from one superhump to the next and amplitudes on a single night varied by up to 0.06 magnitudes.

**Orbital period**

Figure 3(a) shows the power spectrum of the combined photometric data according to the Date Compensated Discrete Fourier Transform (9) (DCDFT), having first subtracted the average magnitude, using the *Peranso* (10) software. We interpret the strongest signal at 12.76(3) cycles/day as being due to the superhumps. The error estimate is derived using the Schwarzenberg-Czerny method (11). The corresponding period of 0.07837(20) d is consistent with our earlier measurement from the times of superhump maximum. We also note the presence of 1 cycle/day aliases of the superhump signal, which is due to the rather short runs and limited longitudinal coverage. There is also a further set of signals at 25.54(3) cycles/day, which represents the second harmonic of the superhump period and its 1 cycle/day aliases.

We attempted to remove the superhump signal by pre-whitening the data with its 12.76 cycles/day period and repeated the analysis. The resulting power spectrum is shown in Figure 3(b) and 3(c). This shows that the superhump signal was not completely removed. The reason for this is that as we showed from the O-C analysis, the superhump period changed during the outburst. The second strongest signal was at 13.28 cycles/day (plus 1 cycles/day aliases), which we tentatively interpret as the orbital signal giving $P_{orb}$ = 0.07531(25) d.

Gaensicke et al. (12) analysed $P_{orb}$ as a function of $P_{sh}$ for 68 SU UMa systems (from an earlier paper by Patterson et al. (13) and found a tight correlation of:

$$P_{orb} \text{ [min]} = 0.9162(52) \times P_{sh} \text{ [min]} + 5.39(52) \qquad \text{Equation 2}$$

Applying this to the superhump period of SDSS J083931.35+282824.0, the "predicted" orbital period is $P_{orb}$ = 108.773 (+/- 1.113) min, or 0.07554(77) d which is consistent with our measurement.

The fractional superhump period excess $\varepsilon = (P_{sh} - P_{orb})/P_{orb}$ is 0.039(6), which we also note is consistent with other SU UMa type dwarf novae of similar $P_{orb}$ listed for example in Patterson *et al.* (13). Measuring $\varepsilon$ provides a way to estimate the mass ratio, $q = M_{sec}/M_{wd}$, of a cataclysmic variable and following Patterson *et al.* (13) we find q ≈ 0.17 for SDSS J083931.35+282824.0.

**Conclusions**

A coordinated CCD photometry campaign was conducted during the first confirmed superoutburst of the recently discovered dwarf nova, SDSS J083931.35+282824.0 during April 2010. The star has an average magnitude of V =19.8 in quiescence and was observed at 14.0 during the outburst, an outburst amplitude of at least 5.8 magnitudes. Only the plateau phase of the outburst was observed during which superhumps with peak-to-peak amplitude of up to 0.28 magnitudes were present, showing SDSS J083931.35+282824.0 to be an SU UMa type dwarf nova. The superhump amplitude gradually decreased during the plateau phase and the star faded at a rate of 0.11 mag/d during this time. Analysis of the times of superhump maximum revealed a mean superhump period of $P_{sh}$ = 0.07836(2) during the first 3 days. This subsequently decreased to 0.07800(3) d. Analysis of the data revealed evidence for an orbital period $P_{orb}$ = 0.07531(25) d. The fractional superhump period excess was ε = 0.039(6), which is consistent with other dwarf novae of similar orbital period. We estimated the mass ratio of the secondary to the primary star, $M_{sec}/M_{wd}$, as q ≈ 0.17 from an empirical relationship between *q* and ε.

**Acknowledgements**

The authors thank Prof. Joe Patterson of the Center for Backyard Astrophysics (CBA) for the encouragement and support he gave to this campaign and Jonathan Kemp (CBA Hilo, Joint Astronomy Centre, Hawaii) for kindly providing the CBA data. We acknowledge the use of data from the Catalina Real-Time Transient Survey. This research made use of SIMBAD and Vizier, operated through the Centre de Donées Astronomiques (Strasbourg, France), and the NASA/Smithsonian Astrophysics Data System. We thank our referees, Prof. Boris Gaensicke and Dr. Chris Lloyd, for helpful comments that have improved the paper.

**Addresses**

JS: "Pemberton", School Lane, Bunbury, Tarporley, Cheshire, CW6 9NR, UK [bunburyobservatory@hotmail.com]

EdM: Departamento de Fisica Aplicada, Facultad de Ciencias Experimentales, Universidad de Huelva, 21071 Huelva, Spain; Center for Backyard Astrophysics, Observatorio del CIECEM, Parque Dunar, Matalascañas, 21760 Almonte, Huelva, Spain [demiguel@uhu.es]

GR: 2007 Cedarmont Dr., Franklin, TN 37067, USA, [georgeroberts@comcast.net]

DFC: Warren Wilson College, Asheville, NC 28815, USA [dcollins@warren-wilson.edu]

GM: 5 Inverness Way, Hillsborough, California 94010-7214, USA [gordonmyers@hotmail.com]

TC: 7021 Whispering Pine, Harrison, AR 72601, USA [jmontecamp@yahoo.com]

| Observer | Telescope | CCD |
|---|---|---|
| de Miguel | 0.25 m reflector | QSI-516ws |
| Campbell | 0.3 m SCT | SBIG ST-8 |
| Collins | 0.2 m SCT | SBIG ST-7X |
| Myers | 0.3 m reflector | FLI 1024 Dream Machine |
| Roberts | 0.4 m SCT | SBIG ST-8 |
| Shears | 0.28 m SCT | Starlight Xpress SXVF-H9 |

**Table 1: Equipment used**

| Start time | End time | Duration (h) | Observer |
|---|---|---|---|
| 2455295.329 | 2455295.449 | 2.9 | Shears |
| 2455295.336 | 2455295.542 | 4.9 | de Miguel |
| 2455296.333 | 2455296.533 | 4.8 | de Miguel |
| 2455296.563 | 2455296.717 | 3.7 | Roberts |
| 2455296.594 | 2455296.753 | 3.8 | Collins |
| 2455297.337 | 2455297.506 | 4.1 | de Miguel |
| 2455297.553 | 2455297.724 | 4.1 | Roberts |
| 2455297.720 | 2455297.845 | 3.0 | Myers |
| 2455298.318 | 2455298.456 | 3.3 | de Miguel |
| 2455298.335 | 2455298.457 | 2.9 | Shears |
| 2455298.549 | 2455298.712 | 3.9 | Roberts |
| 2455298.559 | 2455298.701 | 3.4 | Collins |
| 2455299.628 | 2455299.720 | 2.2 | Campbell |
| 2455301.583 | 2455301.699 | 2.8 | Roberts |

**Table 2: Log of time-series photometry**

| Superhump cycle | Superhump maximum (HJD) | O-C (d) | Error (d) | Superhump amplitude |
|---|---|---|---|---|
| 0 | 2455295.3519 | -0.0005 | 0.0003 | 0.26 |
| 0 | 2455295.3526 | 0.0002 | 0.0003 | 0.28 |
| 1 | 2455295.4303 | -0.0004 | 0.0005 | 0.28 |
| 1 | 2455295.4306 | -0.0001 | 0.0007 | 0.24 |
| 2 | 2455295.5105 | 0.0014 | 0.0011 | 0.21 |
| 13 | 2455296.3719 | 0.0008 | 0.0005 | 0.24 |
| 14 | 2455296.4497 | 0.0003 | 0.0004 | 0.21 |
| 15 | 2455296.5278 | 0.0000 | 0.0006 | 0.24 |
| 16 | 2455296.6065 | 0.0004 | 0.0005 | 0.23 |
| 16 | 2455296.6049 | -0.0012 | 0.0006 | 0.25 |
| 17 | 2455296.6834 | -0.0011 | 0.0006 | 0.19 |
| 17 | 2455296.6837 | -0.0008 | 0.0005 | 0.21 |
| 26 | 2455297.3887 | -0.0010 | 0.0008 | 0.20 |
| 27 | 2455297.4670 | -0.0011 | 0.0008 | 0.20 |
| 29 | 2455297.6265 | 0.0017 | 0.0002 | 0.19 |
| 30 | 2455297.7034 | 0.0002 | 0.0004 | 0.20 |
| 31 | 2455297.7817 | 0.0002 | 0.0013 | 0.21 |
| 38 | 2455298.3314 | 0.0013 | 0.0002 | 0.20 |
| 39 | 2455298.4067 | -0.0017 | 0.0003 | 0.25 |
| 41 | 2455298.5652 | 0.0001 | 0.0002 | 0.24 |
| 41 | 2455298.5683 | 0.0032 | 0.0003 | 0.20 |
| 42 | 2455298.6423 | -0.0012 | 0.0014 | 0.25 |
| 42 | 2455298.6443 | 0.0008 | 0.0013 | 0.19 |
| 55 | 2455299.6582 | -0.0040 | 0.0006 | 0.24 |
| 80 | 2455301.6082 | -0.0130 | 0.0004 | 0.19 |
| 81 | 2455301.6860 | -0.0135 | 0.0019 | 0.16 |

**Table 3: Superhump maximum times and amplitudes**

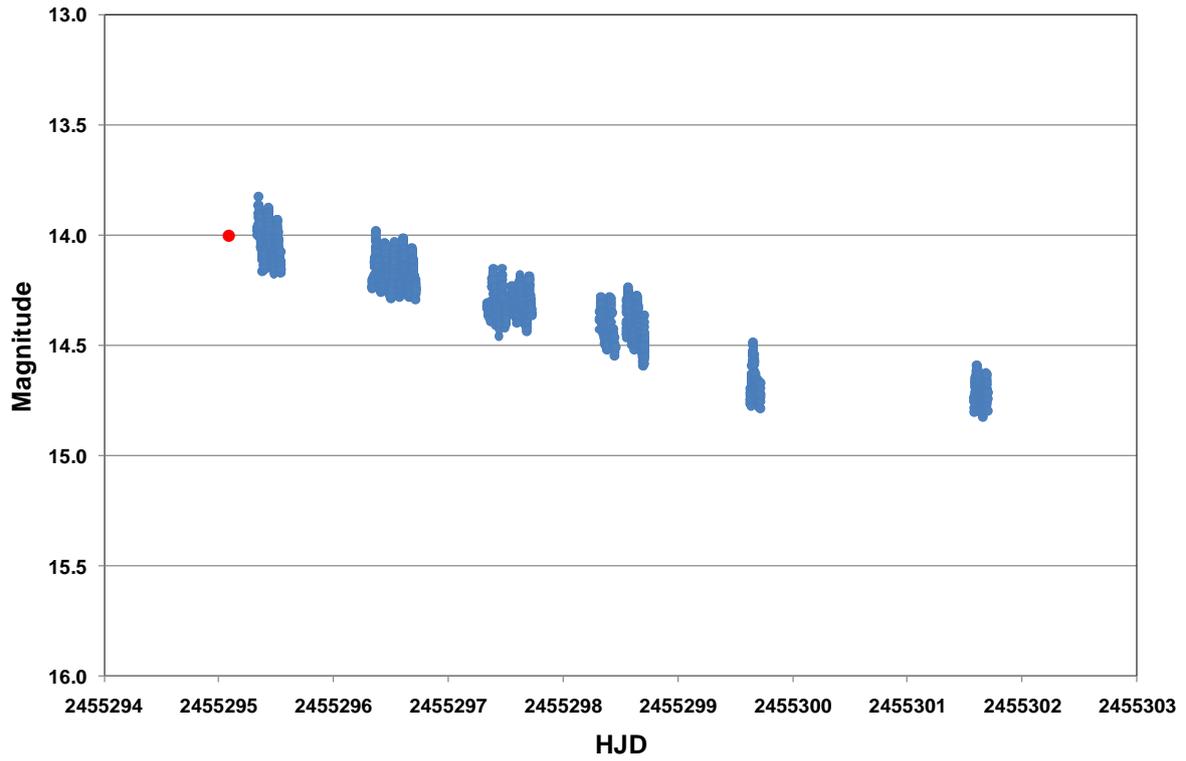

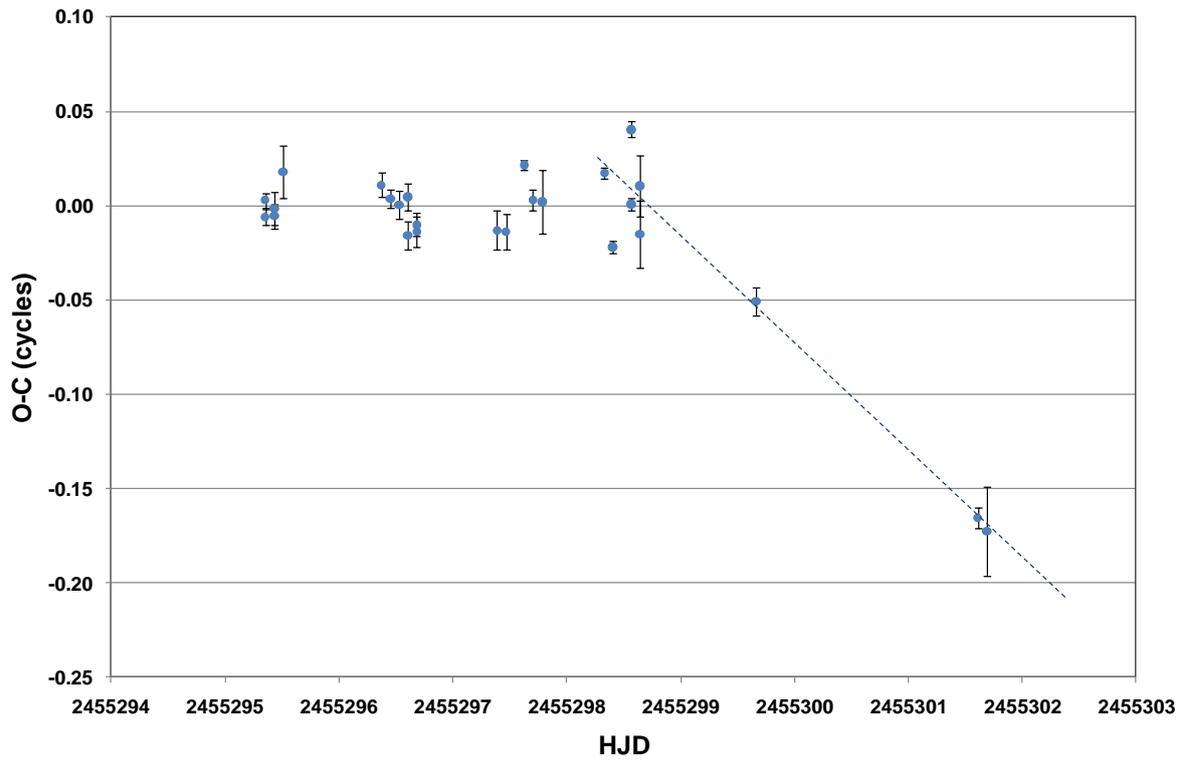

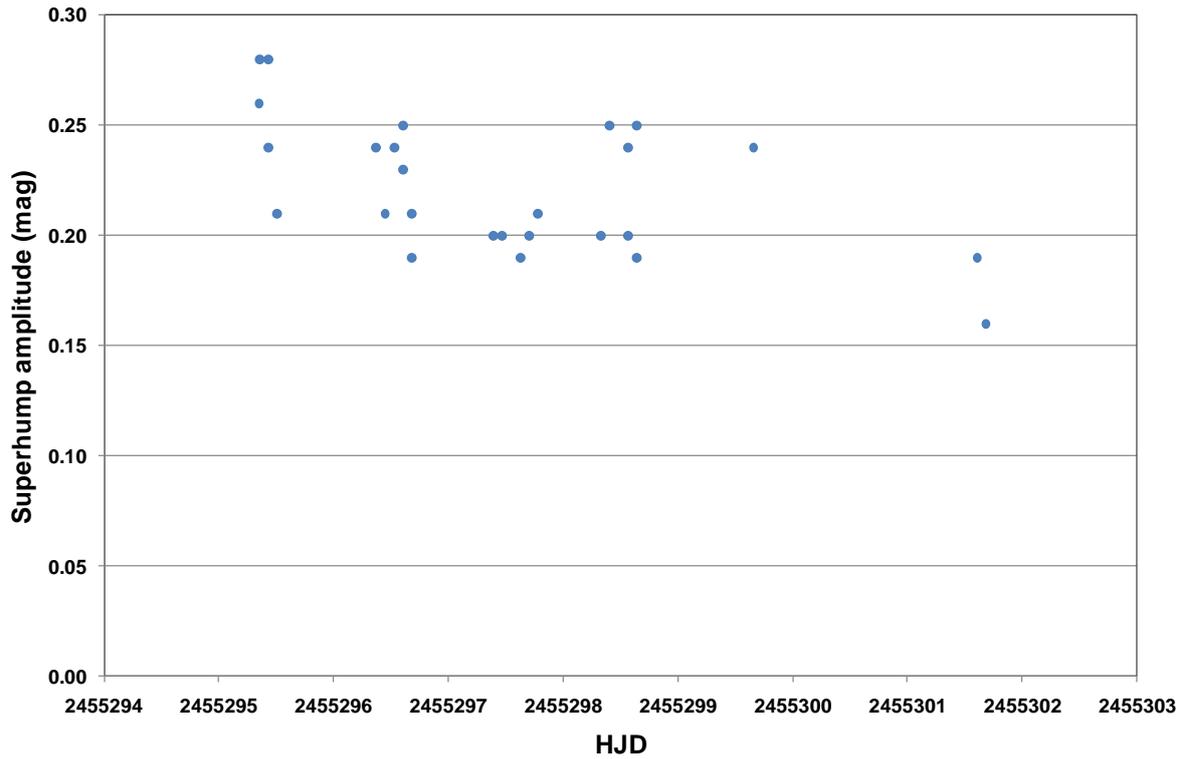

**Figure 1: Light curve of the outburst (top), O-C diagram of superhump maxima relative to the ephemeris in Equation 1 (middle) and superhump amplitude (bottom)**

In the O-C diagram, the blue dotted line is a linear fit to the data between JD 2455298 and 2455301

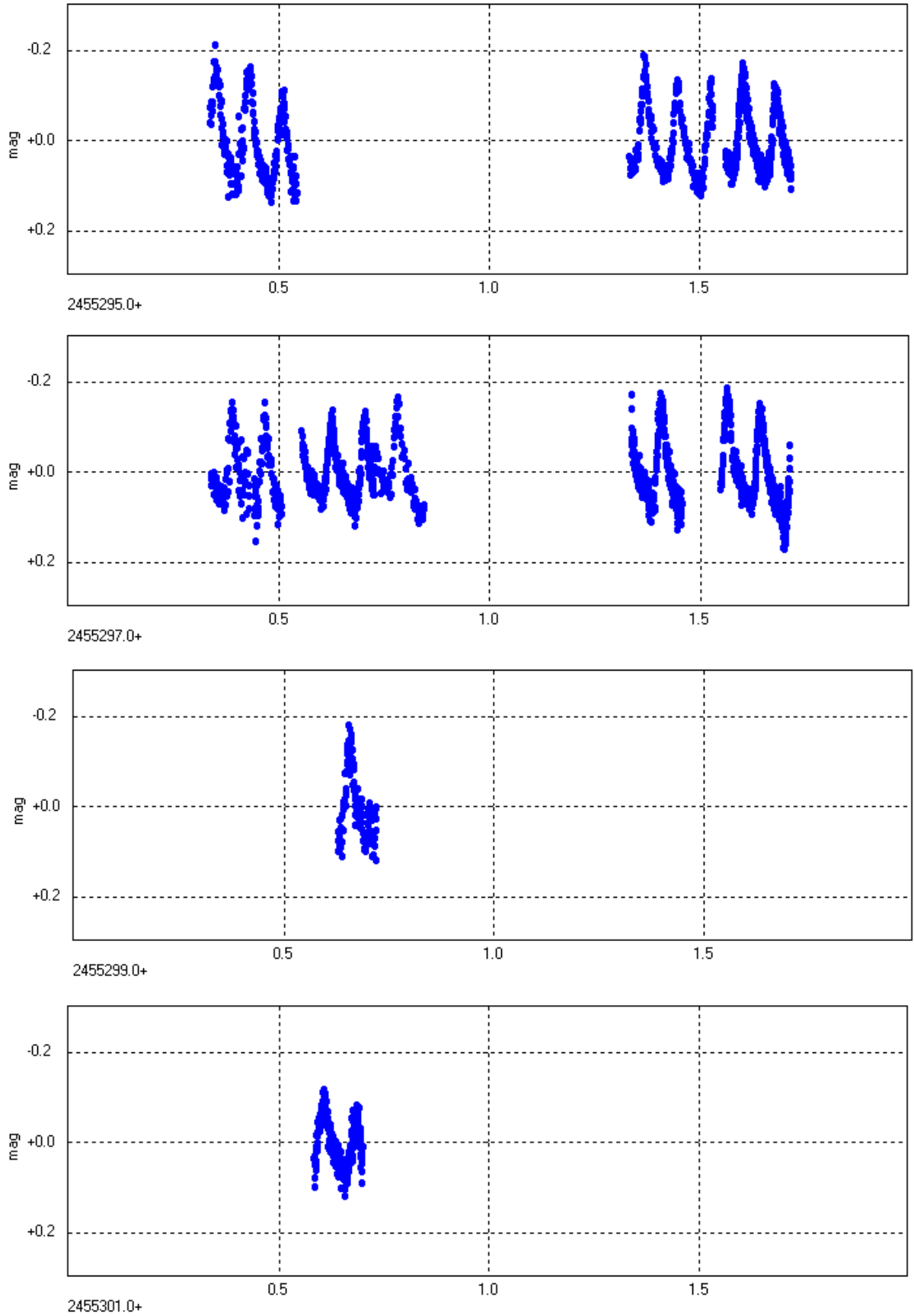

**Figure 2: Time series photometry during the outburst**

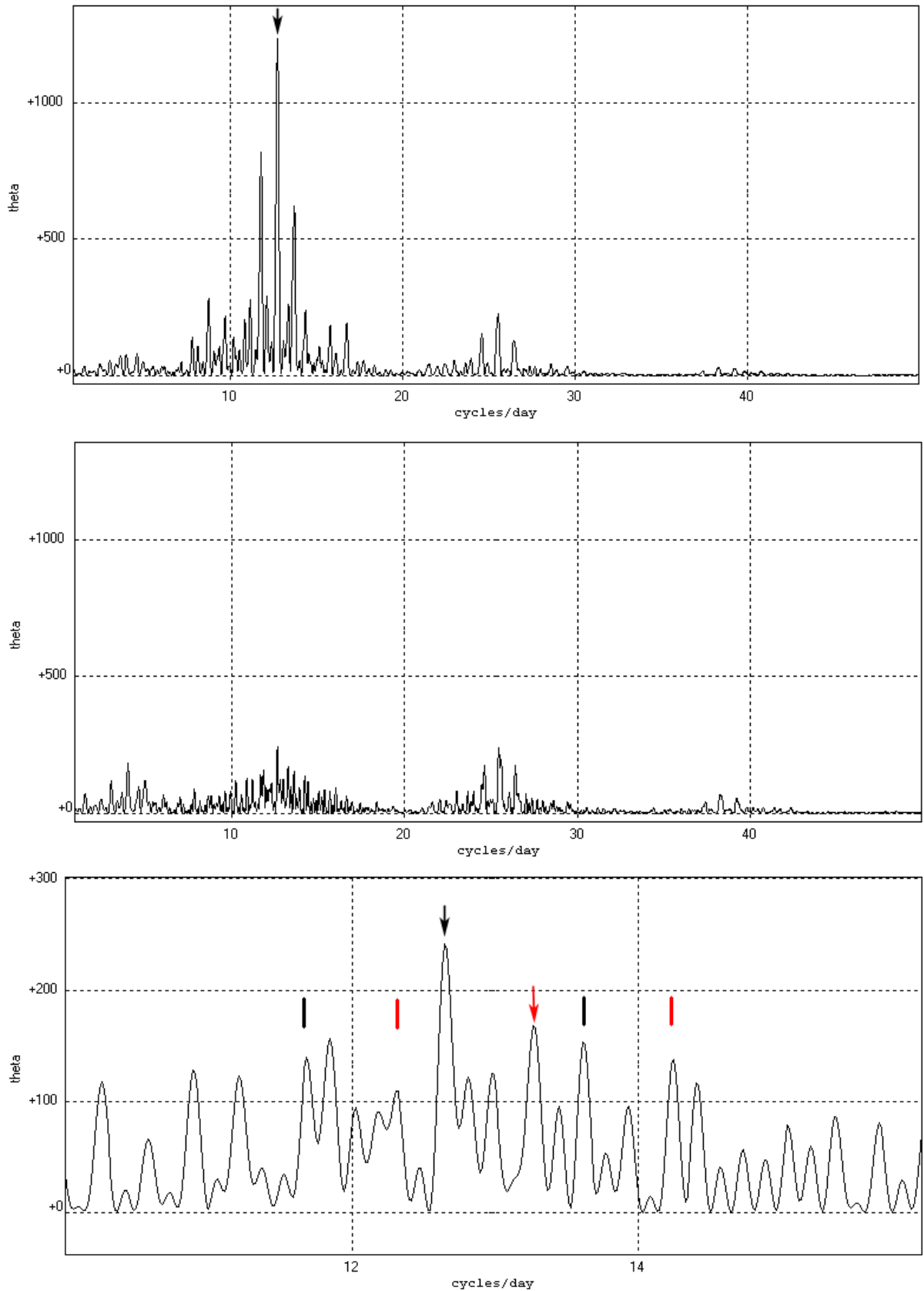

**Figure 3: DCDFT power spectra**
(a) All data, indicating superhump signal at 12.76 cycles/day; (b) Power spectrum after pre-whitening with 12.76 cycles/day; (c) Enlarged detail from (b) indicating the residual superhump signal (black arrow) and its 1 cycles/day aliases (black bars) and the possible orbital signal at 13.28 cycles/day (red arrow) and its 1 cycles/day aliases (red bars)